\documentclass[final,journal]{IEEEtran}

\newif\ifDeepMIMOModel
\DeepMIMOModeltrue

\newif\ifSimpleNParamEq
\SimpleNParamEqtrue

\usepackage{ifpdf}

\ifCLASSINFOpdf
  \usepackage[pdftex]{graphicx}
  \graphicspath{{../pdf/}{../jpeg/}}
  \DeclareGraphicsExtensions{.pdf,.jpeg,.png}
\else
  \usepackage[dvips]{graphicx}
  \graphicspath{{../eps/}}
  \DeclareGraphicsExtensions{.eps}
\fi

\newcommand{\jeremyModif}[1]{#1}

\usepackage{cite}
\usepackage{algorithm}
\usepackage{algorithmic}
\usepackage[cmex10]{amsmath}
\usepackage{amsthm}
\usepackage{amssymb}
\usepackage{balance}
\usepackage{eqparbox}
\usepackage{multirow}

\usepackage{mathtools, nccmath}
\DeclarePairedDelimiter{\nint}\lfloor\rceil

\usepackage{booktabs, siunitx}
\usepackage[svgnames,table]{xcolor}

\usepackage{graphicx}
\usepackage[scaled]{DejaVuSansMono}
\usepackage[T1]{fontenc}

\usepackage{color}
\usepackage{relsize}
\usepackage{mathtools}

\newcommand{\bs}[1]{\boldsymbol{#1}}

\newcommand{\mb}[1]{\mathbf{#1}}
\newcommand{\mr}[1]{\mathrm{#1}}

\newcommand{\bseq}{\begin{subequations}}
\newcommand{\eseq}{\end{subequations}}
\newcommand{\baln}{\begin{align}}
\newcommand{\ealn}{\end{align}}
\newcommand{\balnd}{\begin{aligned}}
\newcommand{\ealnd}{\end{aligned}}
\newcommand{\beq}{\begin{equation}}
\newcommand{\eeq}{\end{equation}}
\newcommand{\beqn}{\begin{eqnarray}}
\newcommand{\eeqn}{\end{eqnarray}}
\newcommand{\beqno}{\begin{eqnarray*}}
\newcommand{\eeqno}{\end{eqnarray*}}
\newcommand{\bma}{\begin{displaymath}}
\newcommand{\ema}{\end{displaymath}}
\newcommand{\bnu}{\begin{enumerate}}
\newcommand{\enu}{\end{enumerate}}
\newcommand{\bce}{\begin{center}}
\newcommand{\ece}{\end{center}}
\newcommand{\btb}{\begin{tabular}}
\newcommand{\etb}{\end{tabular}}
\newcommand{\ba}{\begin{array}}
\newcommand{\ea}{\end{array}}
\newcommand{\change}[1]{#1} 

\begin{document}

\title{RSSI-Based Hybrid Beamforming Design with Deep Learning}
\author{Hamed Hojatian, Vu Nguyen Ha, J\'{e}r\'{e}my Nadal, Jean-Fran\c{c}ois Frigon, and Fran\c{c}ois Leduc-Primeau \\
{\small\'{E}cole Polytechnique de Montr\'{e}al, Montreal, Quebec, Canada, H3T 1J4 \\ 
Emails: \{hamed.hojatian, vu.ha-nguyen, jeremy.nadal, j-f.frigon, francois.leduc-primeau\}@polymtl.ca} }

\maketitle

\begin{abstract}
Hybrid beamforming is a promising technology for 5G millimetre-wave communications. However, its implementation is challenging in practical multiple-input multiple-output (MIMO) systems because non-convex optimization problems have to be solved, introducing additional latency and energy consumption. In addition, the channel-state information (CSI) must be either estimated from pilot signals or fed back through dedicated channels, introducing a large signaling overhead. 
In this paper, a hybrid precoder is designed based only on received signal strength indicator (RSSI) feedback from each user.
A deep learning method is proposed to perform the associated optimization with reasonable complexity.
Results demonstrate that the obtained sum-rates are very close to the ones obtained with full-CSI optimal but complex solutions. Finally, the proposed solution allows to greatly increase the spectral efficiency of the system when compared to existing techniques, as minimal CSI feedback is required.
\end{abstract}


\IEEEpeerreviewmaketitle
\thispagestyle{empty}

\section{Introduction}
Recently, hybrid analog/digital precoding architectures for multiple-input multiple-output (MIMO) systems have drawn considerable interest due to
their ability to achieve high data rates while reducing the number of transmission chains. Therefore, energy-efficient MIMO systems can be designed thanks to such architectures. In the fifth generation of cellular network technology (5G), hybrid beamforming techniques are considered in the millimetre-wave (mmWave) bands \cite{HIX+2015}.
\jeremyModif{Most of these techniques assume that the channel-state information (CSI) is available at the base station (BS). Although several channel estimation techniques for hybrid beamforming have been proposed in the last few years \cite{cheHB,cheHB2}, channel estimation remains a challenging task due to the hybrid structure of the precoding and to the imperfections of the radio-frequency (RF) chain.}


 \jeremyModif{In practical systems, the CSI is generally estimated by exploiting the downlink-uplink channel reciprocity.}
This holds for instance in time-division duplexing and when the user equipment (UE) has the capability to transmit and receive with the same antennas. However, if different antennas are employed, exploiting the channel reciprocity becomes a challenging task. In addition, the frequency bands can be designated for frequency-division duplexing operation due to the unbalanced data transmission between downlink and uplink side in some practical wireless networks. In such cases, channel reciprocity is unavailable. Instead, CSI reporting through dedicated channels is considered, which limits the spectral efficiency of the system.

To avoid high signaling overhead and allow MIMO beamforming systems with high spectral efficiency, we propose a system that does not rely on full CSI feedback or on channel reciprocity. Instead, only the received signal strength indicators (RSSIs) are transmitted by the users. However, designing the hybrid beamforming precoders becomes an even more challenging task. Nevertheless, recent deep learning (DL) techniques have proven to be outstanding tools to deal with such complicated non-convex problems.
Those techniques first train neural networks. This computationally heavy task is generally performed in data centers. Once trained, the neural network is then directly executed in real time at the transmitter (the base station). 
This approach has the potential to reduce the computational complexity and latency of complex signal processing tasks.


DL techniques have been broadly utilized in numerous applications including wireless communications.
Several works have investigated the use of DL to deal with difficult problems within the physical layer, including 
beamforming. 
In \cite{DL_HB}, the authors were inspired by \cite{BCE} to design hybrid precoders with DL techniques at reduced cost and energy consumption. 
Authors in \cite{CSINET} developed a convolution neural network (CNN) called CsiNet to transmit and recover compressed CSI.
Authors in \cite{Framework} consider a multi input single output system and solve several optimization problems under the assumption that the CSI is perfectly known. 
By using DL with fully-connected (FC) layers and decomposition of the CSI in \cite{HBF}, the optimal analog and digital precoders have been estimated. 

In this paper, we propose to apply DL techniques to compute complete precoders for multi-user hybrid beamforming, by exploiting only the RSSI measurements instead of relying on CSI. The use of RSSI measurements reduces the amount of feedback data and thus improves spectral efficiency, while the use of DL allows to compute the precoders with low complexity.
The proposed method jointly predicts the analog and the fully digital beamforming precoders. Then, the hybrid beamforming precoder is deducted.
The results demonstrate that the proposed method can provide sum rates very close to the optimal ones relying on full CSI, while significantly outperforming conventional RSSI-based techniques. 

The   rest   of   the   paper   is   organized   as   follows. Section~\ref{sec:SM} describes the system model and the hybrid beamforming design problem.
The architecture of the proposed deep neural network and the training approach are described in Section~\ref{sec:DNN}. Section~\ref{sec:NR} presents the simulation results and Section~\ref{sec:conclusion} concludes the paper.

\textit{Notation:} Matrices, vectors and scalar quantities are denoted by boldface uppercase, boldface lowercase and normal letters, respectively. The notations $(.)^{\rm{H}}$, $(.)^{\rm T}$ , $|.|$, $\|.\|$ and $(.)^{-1}$ denote Hermitian transpose, transpose, absolute value, $\ell^2$-norm and the matrix inverse, respectively.

\section{Hybrid Beamforming} \label{sec:SM}

\subsection{System Model}

\begin{figure}[t]
	\centering
	\includegraphics[width=85mm]{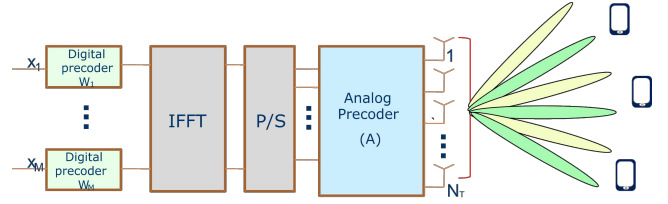}
	\caption{Diagram of a mmWave multi-user system with hybrid analog/digital precoding.}
	\label{fig.HP_system}
\end{figure}

Consider a single-cell multiple-antenna 5G system consisting of a BS equipped with $N_{\sf{T}}$ antennas and $N_{\sf{RF}}$ RF chains serving $M$ single-antenna users. Such system is illustrated in Fig.~\ref{fig.HP_system}.
Simple RSSI measurements feedback and hybrid beamforming (HB) techniques are assumed in this system. 
Let $\mb{w}_m \in \mathbb{C}^{N_{\sf{RF}} \times 1}$ be the digital precoder vector applied to the data symbol $x_{m} \in \mathbb{C}$ intended for user $m$.
Following the precoded digital sequences, the BS then
design an analog precoder (AP) matrix, $\mb{A} = [\mb{a}_1,...,\mb{a}_{N_{\sf{RF}}}] \in \mathbb{C}^{N_{\sf{T}} \times N_{\sf{RF}}}$ to map the RF signals from $N_{\sf{RF}}$ RF chains to $N_{\sf{T}}$ antennas.
In this model, we assume each column of matrix $\mb{A}$, e.g. $\mb{a}_n$, is selected from a set of analog beam codewords $\mathcal{A} = \lbrace \mb{a}^{(1)}, ..., \mb{a}^{(L)} \rbrace$, with $\mathbf{a}^{(i)} \in \{1,-1,i,-i\}^{N_{\sf{T}} \times 1}$ and where $L$ is the length of codebook.
By taking into account of the HB design for the multi-users system in \cite{Alkhateeb-TWC2014}, the signal received by user $m$ can be given by
\begin{equation} \label{recv_sig}
y_m =  \mb{h}_{m}^{\rm{H}} \sum_{\forall j} \mb{A} \mb{w}_j x_j + \eta_m,
\end{equation} 
where $\mb{h}_{m}$ is the channel vector of user $m$ and $\eta_m$ is the additive Gaussian noise at user $m$. 
Assuming coherent detection at the users, the signal-to-interference-plus-noise ratio (SINR) at user $m$ is given by
\begin{eqnarray}\label{SINR}
\mr{SINR}_{m} = \frac{ \big|\mb{h}^{\rm{H}}_{m} \mb{A} \mb{w}_{m} \big|^2}{\sum_{j \neq m} \big|\mb{h}^{\rm{H}} _{m} \mb{A} \mb{w}_{j} \big|^2 + \sigma^2},
\end{eqnarray}
where $\sigma^2$ is the noise power. 
Assuming additive white Gaussian noise (AWGN) channel, the total achievable data-rate of the system is given by 
\beq \label{rate}
R(\mb{W},\mb{A}) = \sum_{\forall m} \log_2(1 + \mr{SINR}_{m}),
\eeq
where $\mb{W}= [\mb{w}_{1}, ..., \mb{w}_{m}]$ is the digital precoder (DP) matrix obtained by concatenating
all the DP vectors $\mb{W}$.
In this paper, we focus on HB design to maximize the network throughput corresponding to the following optimization problem
\begin{subequations}  \label{SEE-max-prb}
\begin{eqnarray} 
&\underset{\left\lbrace \mb{w}_{m} \right\rbrace,\mb{A}}{\max} & R(\mb{W},\mb{A}) = \sum_{\forall m} \log(1 + \mr{SINR}_{m}) \\
& \text{s.t.} & \mb{a}_n \in \mathcal{A}, 1 \leq n \leq N_{\sf{RF}}, \label{cnt_A} \\
&&  \sum_{\forall m} \mb{w}^{\rm{H}} _{m} \mb{A}^{\rm{H}}  \mb{A} \mb{w}_{m} \leq P_{\sf{max}}, \label{cnt2} 
\end{eqnarray}
\end{subequations}
where $P_{\sf{max}}$ stands for the maximum transmission power.

\subsection{Optimal HB Design with Perfect CSI} \label{sec:perf_CSI}
Solving problem \eqref{SEE-max-prb} with perfect CSI has been studied in the literature \cite{Ayach2014,Yu_TSP_16,VuHa_TWC_2018}.
In particular, it has been shown that the HB can be
obtained by employing a framework consisting of two main stages.
In the first stage, the fully digital precoder (FDP) can be obtained by solving the following problem.
\begin{subequations} \label{opt-prob}
\begin{eqnarray} 
&\underset{\left\lbrace \mb{u}_{m} \right\rbrace}{\max} &
\sum_{\forall (m)}  \log\left(1 + \frac{ \big|\mb{h}^{\rm{H}}_{m} \mb{u}_{m} \big|^2}{\sum_{j \neq m} \big|\mb{h}^{\rm{H}} _{m} \mb{u}_{j} \big|^2 + \sigma_m^2} \right)  \\
& \text{ s. t.} & \sum_{\forall m} \mb{u}^{\rm{H}} _{m}\mb{u}_{m} \leq P_{\sf{max}}, \label{cnt_op2}
\end{eqnarray}
\end{subequations}
where $\mb{u}_{m}=\mb{A}{\mb{w}}_{m} $ is the FDP for user $m$ where $\mb{u}_{m} \in \mathbb{C}^{N_{\sf{T}} \times 1}$. A method for solving problem \eqref{opt-prob} can be found in \cite{VuHa_TWC_2018}.
Let $\mb{u}^{\star}_{m}$ be the optimal FDP obtained after solving problem \eqref{opt-prob}.
In the second stage, optimal performance HB, $\mb{w}_{m}$'s and $\mb{A}$, can be re-constructed from the FDP \cite{Ayach2014} via a minimum mean square error (MMSE) approximation as follows.
\beq 
\underset{\left\lbrace \mb{w}_{m} \right\rbrace,\mb{A}}{\min}  \sum_{\forall (m)} \big\|\mb{u}^{\star}_{m} - \mb{A} \mb{w}_{m}\big\|^2  \text{s.t. \eqref{cnt_A}, \eqref{cnt2}}. \label{MMSE_prb}
\eeq
An iterative weighted minimum mean squared error (WMMSE) algorithm can be exploited to reconstruct the HB from FDP. In this approach, one set of DP and AP is optimized alternatingly while keeping the others fixed. For a given AP matrix $\mb{A}$, and assuming that the power constraint (\ref{cnt2}) is temporarily removed, the well-known least squares solution can be obtained as
\beq \label{wpre}
\hat{\mb{w}}_{m} =  \mb{A}^{\dagger} \mb{u}^{\star}_{m}
\eeq
Then, in order to satisfy the power constraints, the DP vector can be normalized as
\beq \label{equ:normalize}
\mb{w}_{m} = \dfrac{\sqrt{P}}{\left( \sum_{\forall m} \big\|\mb{A} \hat{\mb{w}}_{m}\big\|^2\right) ^{1/2}}\hat{\mb{w}}_{m}.
\eeq

While fixing the DP vector $\{\mb{w}_{m}\}$'s, the AP matrix $\mb{A}$ can be optimized by choosing $\mb{A} \in \mathcal{A}$ as follows:
\beq \label{frobenius_prob_A}
\mb{A} = \underset{\mb{A}\in \mathcal{A}}{\arg \min}   \sum \limits_{\forall m} \big\|\mb{u}^{\star}_{m} - \mb{A} \mb{w}_{m}\big\|^2.
\eeq
Then, the complete HB design is summarized in Algorithm~\ref{P7_alg:gms2}.

\begin{algorithm}[!t]
\caption{\textsc{FDP-based HB Design}}
\label{P7_alg:gms2}
\begin{algorithmic}[1]
\STATE Input: Given $\mb{u}^{\star}_{m}$'s.
\STATE Select any $\mb{A}^{(0)} \in \mathcal{A}$, set $l=0$
\REPEAT 
\STATE Fix $\mb{A}^{(l)}$, update $\mb{w}_{m}^{(l)}=\mb{A}^{(l)\dagger}\mb{u}_{m}^{\star}$ for all $m$.
\STATE Fix $\mb{w}_{m}^{(l)}$'s, update $\mb{A}^{(l+1)}$ as in \eqref{frobenius_prob_A}.
\STATE Update $l=l+1$.
\UNTIL Convergence or a stopping criterion trigger.
\STATE Return $\mb{A}^{\star}$ and $\tilde{\mb{w}}_{k,s}^{\star}$'s.
\STATE Normalize $\mb{w}_{m}^{\star}$'s as in (\ref{equ:normalize}).
\end{algorithmic}
\end{algorithm}

However, this optimal method is nearly impossible to implement in real-time systems due to its overwhelming computational complexity. Furthermore, CSI estimation is a difficult task which requires, in practical systems, very large signaling overhead from both transmitter and receiver sides.
For these reasons,  we consider in this paper a novel HB designs based on the codeword RSSI measurements feedback process, a new process applied for multi-user MIMO transmission in the 5G standard, which is discussed in the following section. 

\subsection{Beam Training}
In 5G systems, \emph{beam or codeword training} can be employed to identify the best beam codeword for each UE based on which the HB is designed for all users.
The beam training strategy includes two phases.
In the first phase, the BS transmits $K$ synchronization signal (SS) bursts, each with different analog beam codewords.
Such signals are received and processed by users to establish the initial access (IA), reconnect after beam misalignment, and search for additional BSs for potential handover.
The sounding beams in each SS burst are selected from a predetermined codeword set, which
\jeremyModif{is} intended to facilitate multi-antenna processing in BS and UE as no prior channel information is available during initial accesses.
Then, in the second phase, each user feeds back to the BS the RSSI they computed from the received SS codewords. 
This paper aims to exploit this feedback information for designing the hybrid precoders for all users without estimating the CSI.

We first consider the IA and beam training frame.
Assume that the analog beam codeword $\mb{a}_{\sf{SS}}^{(k)} \in \mathcal{A}$ is selected for broadcasting SS in the $k$-th burst of IA frame.
The sounding signal $\mb{a}_{\sf{SS}}^{(k)} s^{(k)}$ is then broadcast to all users in the network, where $s^{(k)}$ is the synchronization signal for the $k$-th burst. At user $m$, the received synchronization signal can be written as 
\beq
r_{m}^{(k)} = \mb{h}^{(k){\rm{H}} }_m \mb{a}_{\sf{SS}}^{(k)}s^{(k)} + \eta_m^{(k)},
\eeq
where $\mb{h}^{(k)}_m \in \mathbb{C}^{N_{\sf{T}} \times 1}$ stands for the channel vector from the BS to user $m$.
After receiving $r_{m}^{(k)}$, user $m$ estimates the normalized RSSI as
\beq
\alpha_m^{(k)} = \vert r_{m}^{(k)} \vert^2 / \beta,
\eeq
where $\beta$ is a scaling factor that ensures that $\alpha_m^{(k)} \in [0,1] \,\forall k$.
We denote by $\bs{\alpha}_m = [\alpha_m^{(1)}, ..., \alpha_m^{(K)}]^T$ the vector of all RSSI values received at user $m$.
In practical systems, the RSSI vector is quantized. Let $\tilde{\bs{\alpha}}_m$ be the quantized RSSI vector of user $m$ received at BS. Assuming linear quantization, the quantized RSSI vector is given by
\beq
\tilde{\bs{\alpha}}_m = \frac{\nint{\bs{\alpha}_m (2^{N_{b}}-1)}}{(2^{N_{b}}-1)},
\eeq
where $\nint{.}$ is the round operator and $N_{b}$ is the number of quantized bits. 

Without CSI information, designing HB based on $\{ \tilde{\bs{\alpha}}_m \}$ is a challenging problem that it has not been addressed before. 
The following sub-section proposes two simple methods for HB design based on beam training.

\subsection{Simple Beam Training Approaches for HB}\label{sec:HP}
To provide a point of comparison for the DL HB approach, we propose two straightforward methods that can be used to design the FDP using only RSSI values.
\begin{itemize}
\item \textbf{Maximum direction:} The best analog beam codeword in $\{ \mb{a}_{\sf{SS}}^{k} \}$'s selected for each user. In particular, the FDP for user $m$ is estimated as
\beq
\label{MD}
\mb{u}_m^{\sf{max}} = \mb{a}_{\sf{SS}}^{(m_\star)},
\eeq
where $m_\star = \arg \max_{k} \tilde{\alpha}_m^{(k)}$.
\item \textbf{Maximum-ratio combining:} The FDP for user $m$ is selected according to the maximum-ratio combining (MRC) rule as
\beq
\label{MRC}
\mb{u}_m^{\sf{MRC}} = \frac{\sum_{k} \tilde{\alpha}_m^{(k)} \mb{a}_{\sf{SS}}^{(k)}}{\sum_{k} \tilde{\alpha}_m^{(k)}}.
\eeq
\end{itemize}
After estimating the FDP for users, the HB can be obtained using the method described in \ref{sec:perf_CSI}.
However, these methods are sub-optimal and may not provide good performance when considering complex scenarios. Therefore, there is a need to investigate a more sophisticated technique for obtaining the HB from RSSI values.

\section{Deep-Learning Method}\label{sec:DNN}
We consider using deep neural networks (DNN) to predict good hybrid precoders based on the quantized RSSI feedback $\tilde{\bs{\alpha}}$ received at the BS.
Directly predicting the digital part $\mb{W}$ and the analog part $\mb{A}$ of the precoder could lead to a mismatch in the case of imperfect predictions. Instead, we propose to first predict $\mb{A}$ and $\mb{u}_m$ and to compute $\mb{w}_m$ based on \eqref{wpre} and \eqref{equ:normalize}.
Therefore, the proposed method performs the following two steps:
\begin{enumerate}
\item Determine $\mb{A}$ by finding the index of the codeword for every column of $\mb{A}$. Since the elements of $\mb{A}$ are quantized, a multi-label classification network can be designed to predict the index of the analog codeword for each RF chain.
\item To determine the FDP of user $m$, we first express it as a linear combination of the synchronization signal codewords $\mb{a}_{\sf{SS}}^{(k)}$:
\beq \label{upre}
\mb{u}_m = \sum \limits_{k = 1}^K \delta^{k}_m \mb{a}_{\sf{SS}}^{(k)} \, .
\eeq
We then design a DNN to predict the $K$ $\delta^{k}_m$ values.
Since these values are not quantized, we use a regression DNN for this task.
\change{We opt to predict the FDP through the $\delta^{k}_m$ values and \eqref{upre}, rather than predicting $\mb{u}_m$ directly, in order to explicitly control the DNN complexity through parameter $K$. We thus hope that the complexity might scale better as the number of antennas in the system increases. This will be verified in future work.}
\end{enumerate}


\subsection{Proposed DNN Architecture}
\begin{figure}[t!]
	\centering
	\includegraphics[scale=0.56]{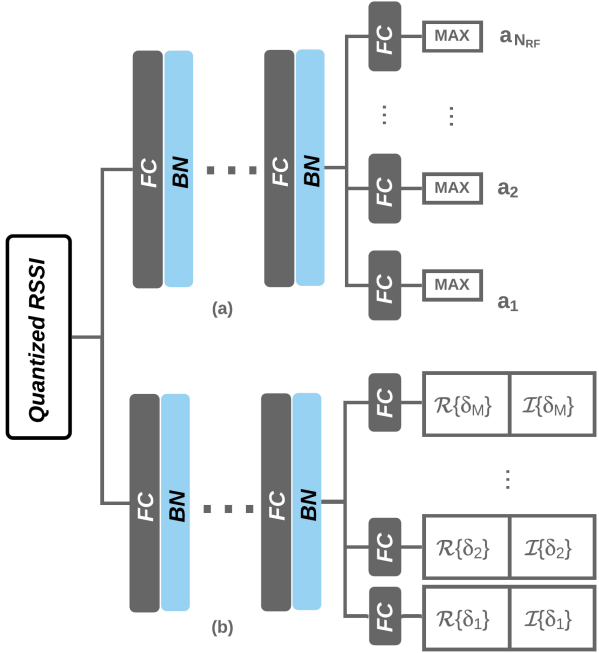}
	\caption{Deep neural network architecture a) Analog-DNN b) $\delta$-DNN}
	\label{fig.DNN}
\end{figure}

As shown in Fig.~\ref{fig.DNN}, two separate DNNs are designed to perform each of the two steps described above. The DNN corresponding to step one is referred to as Analog-DNN and second one is $\delta$-DNN.
The input and hidden layers of both networks have the same architecture, but the output layer of each is adapted to the task at hand. In order to explore the trade-off between performance and complexity, different DNN architectures of varying size are designed, listed in Table~\ref{tbl:DNN}.
We consider two network types: the first is only composed of fully connected layers, and the second also includes two convolutional layers before the FC layers.
For all architectures, the number of neurons in FC layers is fixed to 512, and $N_{f}$ denotes the number of fully connected layers. 
The CNN networks contain two convolutional layers with 128 and 256 channels, respectively, each with convolution kernels of size $3 \times 3$. A $2 \times 2$ \textit{max pooling} operation is applied to the output of the first convolutional layer. 
In both models, batch normalization is applied after each FC layer to reduce overfitting and accelerate the convergence of the training\cite{batchN}.  
Moreover, the leaky rectified linear unit (ReLU) has been used as the activation function\cite{leaky2}. The leaky ReLU function with input $X$ and output $Y$ is given by
\beq
Y =
\begin{cases}
X   &\text{if $X\geq0$}, \\
0.01X &\text{if $X<0$}.
\end{cases}
\eeq
It fixes the \textit{dying ReLU} problem, as it doesn't have zero-slope parts \cite{leakyrelu}, and we observe that it improves accuracy for both the Analog-DNN and $\delta$-DNN compared to a standard ReLU activation. 
In the following sub-sections, both networks are described.

\subsection{Analog-DNN}
The analog precoder is predicted by a multi-label classification network because its targets are quantized. The analog precoder consists of $N_{\sf{RF}}$ codewords. 
Since the BS transmits $K$ analog beam codewords to all $M$ users during the IA, the feedback from all users to the BS is a $K \times M$ matrix, which is used as the input of the DNN. Therefore, the DNN model is trained to predict the optimal codeword for each RF chain according to the RSSI feedback for each user as input. 
For training purposes, the optimal codewords for each RF chain are found by solving \eqref{frobenius_prob_A}. 
We use the conventional cross-entropy loss (CEL) function as the training objective, defined as 
\beq
\mathcal{L}_{CEL} = - \sum^{\Lambda}_{c=1} y_{c} \log(p_{c}),
\eeq
where $\Lambda$ is the number of classes, $y_{c}$ is the true probability for class $c$ and $p_{c}$ is the predicted probability\cite{CEL}. The output of the network gives a probability over all analog codewords. Thus, the final prediction takes the label having the maximum predicted probability. If we assign a label for each individual codeword vectors, we can deploy this loss function to predict the codewords. Consequently, the analog precoder is built by de-mapping the predicted labels.

\begin{table}[t]
\centering
\caption{\vspace{1pt}Parameter selection for the deepMIMO channel model}
\vspace{-4pt}
\begin{tabular}{p{1.2cm}c|p{1cm}c|p{1.3cm}c}
\toprule
\multicolumn{2}{c}{System}  & \multicolumn{2}{c}{Antennas} & \multicolumn{2}{c}{Users} \\
Parameter           &  Value & Parameter   &  Value & Parameter &  Value \\
\midrule
scenario  & ``O1''     & num\_ant\_x & $1$    & active\_BS          & $7$   \\
bandwidth & $0.5$ GHz    & num\_ant\_y & $2$    & act\_user\_first & $1000$  \\
num\_OFDM & $1024$   & num\_ant\_z & $3$    & act\_user\_last  & $1300$  \\
num\_paths& $10$      & ant\_spacing & $0.5$ &  &    \\
\bottomrule
\end{tabular}
\label{tab:deepMIMO_params}
\end{table}

\subsection{$\delta$-DNN}
The $\delta$-DNN model is a regression model that also takes as input the $K \times M$ matrix of RSSI feedback, and outputs the $\delta_m^{(k)}$ values. 
Since the desired outputs are complex-valued, the network outputs the real and imaginary parts separately, for a total number of $2 \times K \times M$ outputs.
Using the DNN output, the FDP is computed using \eqref{upre}, and finally $\mb{w}_{m}$ is deducted from \eqref{wpre} and \eqref{equ:normalize}.

Training is performed using mean squared error (MSE) loss\cite{MSE_Loss}. The MSE loss is defined as 
\beq 
\mathcal{L}_{MSE} =\frac{1}{n}\sum\limits_{i = 1}^n \left(\delta_{i} - \delta ' _{i}\right)^{2},
\eeq
where $\delta_{i}$ is the actual value, $\delta_{i}'$ is the predicted value and $n$ is the number of samples.


\subsection{Dataset Generation}
The deepMIMO channel model\cite{Deepmimo} is employed to generate the train and test dataset. \jeremyModif{In this model, realistic channel information is generated by applying ray-tracing methods to a three-dimensional model of an urban environment. The considered set of channel parameters from this model are summarized in Table~\ref{tab:deepMIMO_params}. The scenario ``O1'' consists of several users being randomly placed in two streets surrounded by buildings. These two streets are orthogonal and intersect in the middle of the map.}
\change{We used supervised training, and} the optimal hybrid precoders required for training can be obtained by employing the methods given in Section~\ref{sec:perf_CSI}.

Additionally, the $\delta^{k}_m$ values used for training can be defined as follows.
\beq
\bs{\delta}^{\sf{opt}}_m = \mb{B}_{\sf{SS}}^{-1} \mb{u}_m^{\sf{opt}},
\eeq
where $\bs{\delta}^{\sf{opt}}_m = [\delta_m^{(1)} ... \delta_m^{(K)}]^T \in \mathbb{C}^{K \times 1}$ , $\mb{B}_{\sf{SS}} = [\mb{a}_{\sf{SS}}^{(1)} ... \mb{a}_{\sf{SS}}^{(K)}]^{T} \in \mathbb{C}^{N_{\sf{T}} \times K}$.
  
\section{Numerical Results}\label{sec:NR}

\begin{figure}[t!]
	\centering
	\includegraphics[width=3.487 in]{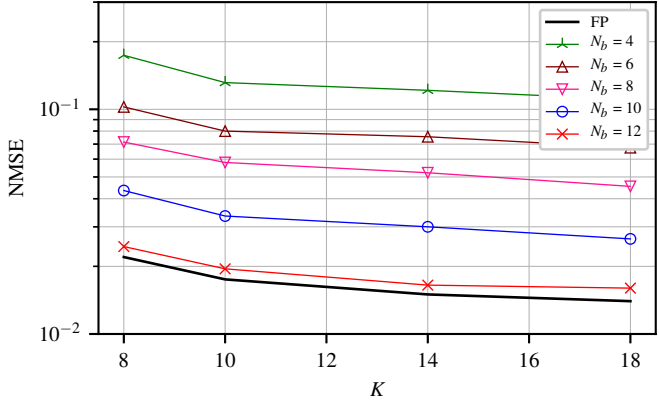}
	\caption{$\delta$-DNN performance versus number of synchronization signals, with different number of quantization bits.}
	\label{fig.NMSERSSI}
\end{figure}

\begin{figure}[t!]
	\centering
	\includegraphics[width=3.487 in]{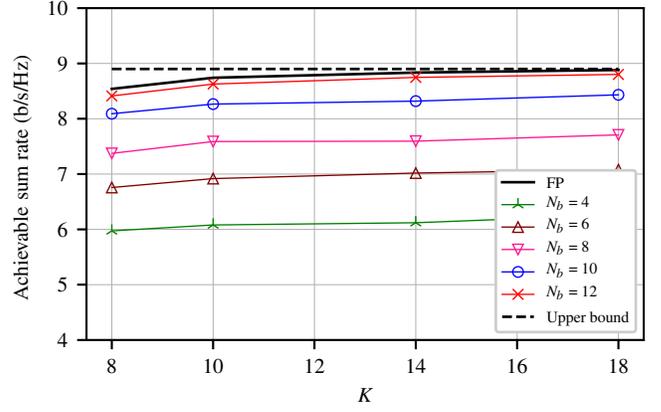}
	\caption{Achievable sum rate (b/s/Hz) versus number of synchronization signals, with different number of quantization bits.}
	\label{fig.RateRSSI}
\end{figure}

In this section, simulations are performed to evaluate the performance of the proposed method. \change{For both $\delta$-DNN and the Analog-DNN networks, \textsc{PyTorch} is employed as the deep learning framework \cite{pytorch}, and the ``Adam'' method with L2 regularization and a weight decay equal to $10^{-6}$ has been used as the optimizer\cite{Adam}. The data set contains a total of 500,000 samples, where $85$\% of the samples are dedicated for the training set and the remaining $15$\% for the test set.} We consider $2$ users communicating with a BS equipped with 6 antennas and 2 RF chains. The noise power is set to $\sigma^2 = 10^{-15}$ W, and the transmit power is normalized to 1 W. The matrix $\mb{a}_{\sf{SS}}$ is chosen such that the rows are mutually orthogonal. When considering the channel attenuation, the SNR ranges from $-5.24$ dB to $28.81$ dB depending on the user positions, with an average of $20.40$ dB if they are randomly positioned.
We consider the FC model with $N_{f} = 7$ for Table~\ref{tbl:sum-rate}, Fig.~\ref{fig.NMSERSSI} and Fig.~\ref{fig.RateRSSI}.

 
%
%

Fig.~\ref{fig.NMSERSSI} shows the normalized mean squared error of the $\delta$-DNN model for different number $K$ of SS bursts per user and different number $N_b$ of quantization bits for the input RSSI values. The NMSE of the $\delta$ weights is given by 
\beq
\mr{NMSE} =  \mathop{\mathbb{E}} \Bigg[ \frac{\|\ \bs{\delta} - \bs{\delta}'\|^2}{\| \bs{\delta} \|^2} \Bigg],
\eeq
where $\|. \|$ represents the Frobenius norm in this context.
Increasing $N_b$ leads to better NMSE performance at the cost of higher feedback signal overhead. With $N_{b} = 12$ bits, the performance of the model is very close to the performance achieved with full-precision (FP) RSSI values. The NMSE performance can also be improved by increasing the number of SS bursts. However, as can be seen in Fig.~\ref{fig.NMSERSSI}, the performance improvement is marginal, and an error floor is observed when $K > 14$. Therefore, the smaller values of $K$ considered are preferable to improve spectral efficiency while reducing the complexity of the DNN.

Fig.~\ref{fig.RateRSSI} compares the achievable sum rate for different values of $K$ and $N_b$. 
The upper bound is the achievable sum rate using the optimum HB with perfect CSI \eqref{opt-prob}. It can be seen that with $N_b = 12$ bits quantization, the sum rate performance is very close to the full precision. By increasing the number of SS bursts, the achievable sum rate is improving, but only marginally, as was observed previously for the NMSE of the $\delta$ values. For this reason, the number of SS bursts will be fixed to $K = 8$ for the remaining results in this section.


\begin{table}[t]
    \centering
    \caption{Sum-Rate Performance}
    \begin{tabular}{lc}
        \toprule
        \multicolumn{1}{c}{} & \multicolumn{1}{c}{Sum rate (b/s/Hz)}\\
        \multicolumn{1}{c}{Method} & \multicolumn{1}{c}{(Ratio w.r.t. optimal HB)} \\
        \cmidrule(lr){1-1} \cmidrule(lr){2-2}         
        {Optimal HB (Perfect CSI) \eqref{opt-prob}}& 8.9\,(100\%) \\
        {\bf Proposed Method (RSSI-Based)} &  $\mathbf{8.54\,(96\%)}$ \\
        {ZFBF (Perfect CSI) \cite{PHS2005}} & 7.38\,$(82\%)$ \\ 
        {Maximum direction (RSSI-Based) \eqref{MD}} & 3.72\,$(41\%)$ \\
        {MRC (RSSI-Based) \eqref{MRC} } & 3.39\,$(38\%)$ \\
        \bottomrule
    \end{tabular}
    \label{tbl:sum-rate}
\end{table}

Table~\ref{tbl:sum-rate} illustrates the sum-rate performance of the proposed method and the sum rate obtained through the following BF precoders: the optimal HB method, the zero-forcing beamforming (ZFBF) \cite{PHS2005}, the maximum direction and the MRC methods presented in Section~\ref{sec:HP}. The RSSIs are assumed to be received with full precision. For the optimal hybrid beamforming and the ZFBF cases, we assume the CSI is perfectly known. The results show that the proposed method outperforms the two others RSSI-based technique. Particularly, the sum rate achieved by the proposed DNN is $3$ times the sum rate obtained with the simple maximum direction technique. Furthermore, a sum-rate loss of only $8$\% is observed between the proposed technique and the optimal one, the later being advantaged since it uses perfect CSI.

Table~\ref{tbl:DNN} compares the accuracy of the Analog-DNN model when considering the FC and the CNN networks with different number of layers. For the Analog-DNN, the accuracy is defined as the number of correct predictions divided by the total number of samples, where a prediction is considered \emph{correct} if it is identical to the optimal codeword.
It is interesting to note that although the accuracy of the Analog-DNN is not excellent, a near optimal sum rate can be still be reached. This can be explained by the fact that the model often outputs a sub-optimal solution that is close to the optimal one. Moreover, the $\delta$-DNN model achieves excellent NMSE performance, which has a positive impact on the obtained sum rate.

The complexity of each network architecture is evaluated based on their number of parameters for both $\delta$-DNN and Analog-DNN.
\ifSimpleNParamEq
\jeremyModif{The number of parameters for a given layer is $\mathcal{P} = N_iN_o W^2+ N_o$, where $N_i$ is the input numbers of the layer, $N_o$ its output numbers and $W$ the kernal size for a CNN layer ($W = 1$ for a FC layer). }
\fi
%
\begin{table}[t]
    \centering
    \caption{Number of parameters and accuracy for $\delta$-DNN and Analog-DNN}
    \begin{tabular}{
        ccccc}
        \toprule
        \multicolumn{3}{c}{} & \multicolumn{1}{c}{$\delta$-DNN} &
        \multicolumn{1}{c}{Analog-DNN}\\
        \; Network& $N_{f}$ & {\# Parameters} & {Accuracy (NMSE)} & {Accuracy} \\
        {Architecture} & {}& {($\times 10^6$)} & {({$\times 10^{-2}$})} & {(\%)} \\
        \midrule
        FC & 5         & 5.8  & 3.2 & 62.1  \\
        \textbf{FC} & {\textbf{7}}      & $\mathbf{6.8}$ & $\mathbf{2.2}$ & $\mathbf{63.9}$  \\
        FC & 9         &  7.9 & 2.1 &  64.4  \\
        CNN & 3 &  12.7 & 3.8 & 64.2  \\
        CNN & 5 &  13.7 & 2.9 & 64.8  \\
        \bottomrule
    \end{tabular}
    \label{tbl:DNN}
\end{table}
 An excessive number of parameters can lead to costly and energy-consuming hardware implementations. Therefore, a trade-off between complexity and performance has to be evaluated.  
 Among the $5$ network configurations evaluated, we conclude that the FC with $N_f=7$ layers offers the best performance-complexity trade-off. 

\section{Conclusion}\label{sec:conclusion}

Hybrid beamforming is a key enabler for millimetre-wave communication. However, the design of digital and analog precoders is quite challenging, and the estimation of the CSI introduces important signaling overhead. In this paper, we propose to design the hybrid beamforming by using deep-learning methods. The novelty resides in the fact that the precoders are generated from the RSSI feedback transmitted by each user. Therefore, no CSI feedback is required, improving the spectral efficiency of the communication system. Finally, we evaluated the performance of the proposed algorithm on the realistic \emph{deepMIMO} channel model. The results demonstrate that, despite not having the CSI, the BS can be trained to design the hybrid beamforming while achieving near-optimal sum rates. Moreover, the proposed method can be implemented in real-time systems thanks to its low computational complexity.

\section*{Acknowledgement}
The authors would like to thank The Institute for Data Valorisation (IVADO) for its support.

\bibliographystyle{IEEEtran}
%
\bibliography{bib/VuHa_5G}

\end{document}